\documentclass [aps,amsmath,amssymb,superscriptaddress,preprint]{revtex4}
\usepackage{graphicx}
\usepackage{xcolor}

\def\gsim{\mathrel{\rlap{\lower4pt\hbox{\hskip1pt$\sim$}}
    \raise1pt\hbox{$>$}}}       %greater than or approx. symbol

\newcommand{\gn}{G_{\rm N}}
\renewcommand{\d}{\mbox{${\rm d}$}}
\newcommand{\rs}{R_{\rm s}}

\begin{document}
\title{Quantum Hair from Gravity}
\author{Xavier Calmet} \email{x.calmet@sussex.ac.uk}
\affiliation{Department of Physics and Astronomy,\\
University of Sussex, Brighton, BN1 9QH, United Kingdom}
\author{Roberto Casadio} \email{casadio@bo.infn.it}
\affiliation{Dipartimento di Fisica e Astronomia, Universit\`a di Bologna, via Irnerio 46, I-40126 Bologna, Italy}
\affiliation{
I.N.F.N., Sezione di Bologna, IS - FLAG, via B.~Pichat~6/2, I-40127 Bologna, Italy
}
\author{Stephen~D.~H.~Hsu} \email{hsusteve@gmail.com}
\affiliation{Department of Physics and Astronomy\\ Michigan State University \\}
\author{Folkert Kuipers}\email{f.kuipers@sussex.ac.uk}
\affiliation{Department of Physics and Astronomy,\\
University of Sussex, Brighton, BN1 9QH, United Kingdom}

\begin{abstract}
We explore the relationship between the quantum state of a compact matter source and of its asymptotic graviton field. For a matter source in an energy eigenstate, the graviton state is determined at leading order by the energy eigenvalue. Insofar as there are no accidental energy degeneracies there is a one to one map between graviton states on the boundary of spacetime and the matter source states. Effective field theory allows us to compute a purely quantum gravitational effect which causes the subleading asymptotic behavior of the graviton state to depend on the internal structure of the source. This establishes the existence of ubiquitous quantum hair due to gravitational effects.
\end{abstract}

\maketitle
\section{Introduction}

Classical no-hair theorems limit the information that can be obtained about the internal state of a black hole by outside observers \cite{Misner:1973prb}. External features (``hair'') of black hole solutions in general relativity are determined by specific conserved quantities such as mass, angular momentum, and charge. In this letter we investigate how the situation changes when both the matter source (black hole interior state) and the gravitational field itself are quantized. The idea of quantum hair has been discussed in various approaches to quantum gravity, see e.g. \cite{Giddings:1993de,Moss:1992ne,Preskill:1990ty,Coleman:1992rn}.

We begin by showing that the graviton state associated with an energy eigenstate source is determined, at leading order, by the energy eigenvalue of the source. These graviton states can be expressed as coherent states of non-propagating graviton modes, with explicit dependence on the source energy eigenvalue. Semiclassical matter sources (e.g., a star or black hole) are superpositions of energy eigenstates with support in some band of energies, and produce graviton states that are superpositions of the coherent states. Next, we
use effective field theory to compute $r^{-3}$ and $r^{-5}$ corrections to the $r^{-1}$ Newtonian potential. These corrections originate from non-local terms in the quantum corrections to Einstein's equations.

We show that the $r^{-5}$ corrections are sensitive to the internal structure of the matter source. That is, two matter sources with the same semiclassical mass $M$ can produce different $r^{-5}$ terms in the metric. These observations imply that information about the interior state of a black hole exists outside the classical horizon. This could, in principle, affect the Hawking radiation states produced as the hole evaporates. We discuss implications for black hole information and holography in the conclusions.

\section{Asymptotic quantum states of the graviton field}
General relativity relates the spacetime metric to the energy-momentum distribution of matter, but only applies when both the metric (equivalently, the gravitational field) and matter sources are semiclassical. A theory of quantum gravity is necessary to relate the quantum state of the gravitational field to the quantum state of the matter source.

A semiclassical matter configuration $S$ is a superposition of energy eigenstates with support concentrated in some narrow band of energies
\begin{equation}
\psi_S = \sum_n c_n \psi_n~~,   \end{equation}    
where $\psi_n$ are energy eigenstates with eigenvalues $E_n$.
$S$ produces a gravitational field (metric) governed by the Einstein equations
$G_{\mu \nu} = 8 \pi \,G_{\rm N}\, T_{\mu \nu}$, 
where the energy momentum tensor is itself semiclassical. Here we assume that $S$ is compact -- localized in some spatial region of an otherwise empty universe -- and consider the gravitational field asymptotically far away.

What can be said about the quantum state of the graviton field given the exact quantum state of the matter source? This question extends beyond the realm of classical general relativity, but we show below that the properties of semiclassical gravity constrain the result in an interesting way.

We find that the quantum state of the asymptotic gravitational field of a matter source which is an energy eigenstate is controlled by the energy eigenvalue $E_n$. In particular, energy eigenstate sources with different eigenvalues produce distinct graviton states. This immediately implies that the asymptotic graviton field of a typical semiclassical matter source is a superposition state of the form
\begin{equation}
\label{super}
\psi_g (S) = \sum_n c_n \psi_g (E_n)   ~~, 
\end{equation}
where $\psi_g (E) \neq \psi_g (E')$ when $E \neq E'$.

It is typically assumed in many body physics that there are no accidental degeneracies -- i.e., that the eigenvalues $E_n$ of a complex matter system are distinct (barring exact symmetries of the Hamiltonian; note even these may be violated by quantum gravity effects), although energy level splittings might be exponentially small in the size of the system. If this is the case, then the above results imply that the state of the matter system can, in principle, be reconstructed from the asymptotic graviton state. The quantum information encoded in the matter system is also stored, via entanglement, in the spacetime metric at infinity.

%\bigskip
%\bigskip

To obtain the desired result we use the following gedanken construction. In brief, we want to show that the matter source energy eigenstate $\psi_n$ produces a different asymptotic graviton state than another state $\psi_{n'}$ of the system with $E_{n'} \neq E_{n}$. The problem is that the energy splitting $E_{n'} - E_n$ could be exponentially small in the size of $S$ and as far as the classical Einstein equations are concerned the corresponding sources $T$ and $T'$ are effectively identical. 

However, we can imagine configurations made of $N$ identical copies of the original system $S$, which we take to be an energy eigenstate ($\psi_S=\psi_n$), and the same number $N$ of identical copies of the system $S'$ with the source in the eigenstate $\psi_{S'}=\psi_{n'}$.
For sufficiently large $N$ the difference in source terms $T$ and $T'$ becomes macroscopic, and the difference between the corresponding metrics is governed by the classical Einstein equations. The asymptotic behaviors of these metrics are equivalent to the additive Newtonian gravitational potentials resulting from each of the $N$ copies of $S$ and $S'$, respectively. Hence the asymptotic graviton state $\psi_g (N E_n)$ of the system $S$ cannot be identical to $\psi_g (N E_{n'})$ of the system $S'$, otherwise the resulting sums would also be identical. See \cite{FN} for additional details of this construction.

%\smallskip

This analysis does not determine the graviton states $\psi_g (E)$, but does establish that different energies $E$ correspond to different (albeit possibly very similar) states $\psi_g$.

%\bigskip

We can obtain the same result via quantum field theory using the property that the spin-2 graviton $h_{\mu \nu}$ couples to the operator $T_{\mu \nu}$. The gravitational potential is generated by graviton exchange between the source ``particle'' $S$ and a test mass. At long wavelengths, we can treat the composite state $S$ as a single particle, analogous to a nucleon which is composite and has its own complex substructure. The Feynman amplitude for graviton emission from an incoming source particle $S$ has a vertex factor which is simply its energy eigenvalue $E$. States $S$ with different energies $E$ have different graviton emission amplitudes, and hence produce different asymptotic states of the $h_{\mu \nu}$ field.

%\bigskip

The graviton quantum state $\psi_g (E)$ is exactly analogous to the quantum state of the $U(1)$ vector field (Coulomb potential) created by a charge $Q$ \cite{Barnich,Mueck,Casadio:2021eio}. This can be constructed explicitly as a coherent state
\begin{equation}
\vert 0 \rangle_Q = \exp \, \left[ Q \, \int d^3k \, q(k) b^{\dagger}(k) \right] \, \vert 0 \rangle_{Q=0}    \label{coherent}
\end{equation}    
where $b^{\dagger}(k)$ is a linear combination of creation operators of the non-propagating (temporal and longitudinal, depending on choice of gauge) modes of the photon. The factor of $Q$ in the exponent shows how the photon state depends on the source charge. In the gravitational case $Q$ is replaced by the energy eigenvalue of the source state and the coherent state modes are temporal and longitudinal graviton modes. In both gauge theory and gravity the manner in which the charge or energy control the asymptotic quantum state is determined by the Gauss law via constrained quantization. Note (\ref{coherent}) is a formal expression which avoids some infrared issues: the $q(k)$ are not normalizable with respect to the Lorentz-invariant 1-particle norm.

The direct connection between the gravitational field (Schwarzschild metric) and the Coulomb potential can also be seen as a consequence of the double copy relationship \cite{oconnell}. For our purposes the most important point is that $\psi_g (E)$ depends explicitly on $E$ and for each distinct energy eigenstate of the compact source there is a different graviton quantum state.

The evaporation of a black hole takes place over a timescale $\sim M^3$ so its evolution from a matter configuration to outgoing radiation is confined to a finite region of spacetime. Hence the asymptotic gravitational field at $r \gg M^3$ remains unchanged, in the form (\ref{super}), throughout the entire process. However, near the horizon the gravitational quantum state presumably reflects the changing internal state of the hole. The internal state is itself dependent on the previously emitted Hawking radiation -- e.g., due to conservation of energy, angular momentum, etc. This provides a mechanism connecting the region just outside the horizon, where the {\it next} quantum of Hawking radiation originates, to the internal state of the black hole and the radiation quanta emitted in the past. Once we go beyond the semiclassical approximation the amplitude for radiation emission is a function of $\psi_g (E)$ which itself depends on the internal state of the hole. We discuss this further in the conclusions.

\section{Leading corrections from quantum gravity}

In general relativity, Birkhoff's theorem states that any spherically symmetric solution of the vacuum field equations must be static and asymptotically flat. In other words, the exterior solution must be given by the Schwarzschild metric. It has been shown that this is not the case in quantum gravity \cite{Calmet:2017qqa,Calmet:2019eof}: the asymptotic gravitational potential of a compact object received quantum gravitational corrections \cite{Calmet:2019eof,Calmet:2020tlj} which are not present for an eternal black hole \cite{Calmet:2019eof,Calmet:2018elv}. Quantum gravitational corrections depend on the composition of the compact object. This quantum memory effect has also been observed in FLRW cosmology \cite{Donoghue:2014yha}. In this section, we show that compact objects are hairy in quantum gravity. We work within the framework of the effective quantum gravitational action at second order in curvature ~\cite{Weinberg:1980gg,Barvinsky:1983vpp,Barvinsky:1985an,Barvinsky:1987uw,Barvinsky:1990up,Buchbinder:1992rb,Donoghue:1994dn,Calmet:2018elv}: 
$\Gamma[g] = \Gamma_{\rm L}[g] + \Gamma_{\rm NL}[g]$, where the local part of the action is given by
\begin{eqnarray}
\Gamma_{\rm L}
=
\int \d^4 x\, \sqrt{g} 
\left[ \frac{{\cal R}}{16\,\pi\, \gn}
+
c_1(\mu)\, {\cal R}^2
+ c_2(\mu)\, {\cal R}_{\mu\nu}\, {\cal R}^{\mu\nu}
+ c_3(\mu)\, {\cal R}_{\mu\nu\alpha\beta}\,
{\cal R}^{\mu\nu\alpha\beta}
\right]
\end{eqnarray}
and the non-local part of the action by
\begin{eqnarray}
\Gamma_{\rm NL}
=
-\int \d^4 x\, \sqrt{g}
\left[
\alpha\, {\cal R}\, \ln{\!\left(\frac{\Box}{\mu^2}\right)} \, {\cal R}
+
\beta\, {\cal R}_{\mu\nu} \, \ln\!{\left(\frac{\Box}{\mu^2}\right)} \, {\cal R}^{\mu\nu}
+
\gamma \, {\cal R}_{\mu\nu\alpha\beta} \, \ln\!{\left(\frac{\Box}{\mu^2}\right)} \, {\cal R}^{\mu\nu\alpha\beta}
\right]
\ .
\label{eq:NLterms}
\end{eqnarray}
This effective action is obtained by integrating out the fluctuations of the graviton and potentially
other massless matter fields. The Wilson coefficients of the local part of the action are not calculable from first principles, as we do not specify the ultra-violet theory of quantum gravity. However, those of the non-local part are calculable and model independent quantum gravitational predictions. These non-local coefficients can be found in e.g. \cite{Calmet:2019eof}.
The equations of motion obtained from varying the effective action with respect to the metric are given by
\begin{eqnarray}
{\cal R}_{\mu\nu} - \frac{1}{2}\, {\cal R}\, g_{\mu\nu} + 16\,\pi\,\gn \left( H_{\mu\nu}^{\rm L} + H_{\mu\nu}^{\rm NL} \right)
= 8 \, \pi \, G_N T_{\mu\nu}
\ .
\end{eqnarray}
The local part of the equation of motion is given by
\begin{align}
H_{\mu\nu}^{\rm L} 
=
&\,  
\bar{c}_1
\left( 2\, {\cal R}\, {\cal R}_{\mu\nu} - \frac{1}{2}\, g_{\mu\nu}\, {\cal R}^2 + 2\, g_{\mu\nu}\, \Box {\cal R} - 2 \nabla_\mu \nabla_\nu {\cal R}\right) 
\label{eq:EQMLoc}
\\
&\,
+\bar{c}_2
\left( 2\, {\cal R}_{~\mu}^\alpha\, {\cal R}_{\nu\alpha} - \frac{1}{2}\, g_{\mu\nu}\, {\cal R}_{\alpha\beta}\, {\cal R}^{\alpha\beta}
+ \Box {\cal R}_{\mu\nu} + \frac{1}{2}\, g_{\mu\nu}\, \Box {\cal R}
- \nabla_\alpha \nabla_\mu {\cal R}_{~\nu}^\alpha
- \nabla_\alpha \nabla_\nu {\cal R}_{~\mu}^\alpha \right)
\ ,
\nonumber
\end{align}
with $\bar{c}_1 = c_1 - c_3$ and $\bar{c}_2 = c_2 + 4\, c_3$.
Finally, the non-local part reads
\begin{align}
H_{\mu\nu}^{\rm NL} 
=
&\,
 - 2\,\alpha
 \left( {\cal R}_{\mu\nu} - \frac{1}{4}\, g_{\mu\nu}\, {\cal R}
 + g_{\mu\nu}\, \Box
 - \nabla_\mu \nabla_\nu \right)
 \ln\left(\frac{\Box}{\mu^2}\right)\, {\cal R}
 \nonumber
 \\
&\, 
- \beta
\bigg( 2\, \delta_{(\mu}^\alpha\, {\cal R}_{\nu)\beta}
- \frac{1}{2}\, g_{\mu\nu}\, {\cal R}_{~\beta}^\alpha
+ \delta_{\mu}^\alpha\, g_{\nu\beta}\, \Box
+ g_{\mu\nu}\, \nabla^\alpha \nabla_\beta  \nonumber  \\ \nonumber 
&\quad 
- \delta_\mu^\alpha\, \nabla_\beta \nabla_\nu
- \delta_\nu^\alpha\, \nabla_\beta \nabla_\mu 
\bigg)
\ln\left(\frac{\Box}{\mu^2}\right)\, {\cal R}_{~\alpha}^\beta &
\nonumber
\\
&\,
- 2 \,\gamma
\left(
\delta_{(\mu}^\alpha\, {\cal R}_{\nu)~\sigma\tau}^{~\beta}
- \frac{1}{4}\, g_{\mu\nu}\, {\cal R}^{\alpha\beta}_{~~\sigma\tau}
+\left( \delta_\mu^\alpha\, g_{\nu\sigma} + \delta_\nu^\alpha\, g_{\mu\sigma} \right)
\nabla^\beta \nabla_\tau \right)
\ln\left(\frac{\Box}{\mu^2}\right)\, {\cal R}_{\alpha\beta}^{~~\sigma\tau}
\ .
\end{align}
Note that the variation of the $\ln \Box$ term yields terms of higher order in curvature
and can thus safely be ignored at second order in curvature. The non-local parts of the field equations are responsible for the memory effect. We can easily illustrate this by considering the corrections to the metric of a stationary homogeneous and isotropic star with radius $\rs$ and density 
\begin{equation}
\rho(r)
=
\rho_0\,
\Theta(\rs-r)
=
\begin{cases}
\rho_0
&
{\rm if}\ r<\rs
\\
0
&
{\rm if}\ r>\rs
\ ,
\end{cases}
\label{rho0}
\end{equation}
where $\rho_0>0$ is a constant and $\Theta(x)$ is Heaviside's step function.
The solution to the Einstein equation inside this star (for $r\le\rs$) is the well-known interior Schwarzschild metric
\begin{align}
\label{eq:IntSchw}
\d s^2
&=
\left( 3\, \sqrt{1 - \frac{2\, \gn\, M}{\rs}} - \sqrt{1 - \frac{2\, \gn\, M \,r^2}{\rs^3}} \right)^2 
\frac{\d t^2}{4} 
-
\left(1 - \frac{2\, \gn\, M r^2}{\rs^3}\right)^{-1}\d r^2
-
r^2\,\d\Omega^2
\nonumber
\\
&\equiv
g_{\mu\nu}^{\rm int}\,\d x^\mu\,\d x^\nu
\ ,
\end{align}
where
\begin{equation}
M
=
4\,\pi
\int_0^{\rs}
\rho\,r^2\,\d r
=
\frac{4\,\pi}{3}\,\rs^3\,\rho_0
\end{equation}
is the total Misner-Sharp mass of the source. 
The corresponding pressure is of order $\gn$ \cite{Calmet:2019eof} in agreement with the fact that the pressure does not gravitate in
Newtonian physics.
Of course, the metric outside the star (for $r>\rs$) is the usual vacuum Schwarzschild
metric
\begin{equation}
\label{eq:ExtSchw}
\d s^2 =
\left(1 - \frac{2\,\gn\, M}{r}\right) \d t^2
-
\left(1 - \frac{2\,\gn\, M}{r}\right)^{-1}
\d r^2
- r^2\,\d\Omega^2
\equiv
g_{\mu\nu}^{\rm ext}\,\d x^\mu\,\d x^\nu
\ ,
\end{equation}
from which one can see that $M$ is also the Arnowitt-Deser-Misner (ADM) mass of the system.
\par

We now perturb the above metrics: $\tilde{g}_{\mu\nu} = g_{\mu\nu} + g_{\mu\nu}^{\rm q}$,
and take the perturbation $g_{\mu\nu}^{\rm q}$ to be $\mathcal{O}(\gn)$.
We solve this equation, imposing the solution to be spherically symmetric and time independent.
In addition we fix the gauge freedom by setting $g_{\theta\theta}^{\rm q}=0$.
Doing so, we obtain the quantum corrections $g_{\mu\nu}^{\rm q}=\delta g_{\mu\nu}^{\rm ext}$ to the
Schwarzschild metric~\eqref{eq:ExtSchw} outside the star. The corrections are
given in \cite{Calmet:2019eof}:
\begin{align}
\label{eq:MetricCorr}
	\delta g_{tt}^{\rm ext}
	&=
	(\alpha + \beta + 3\,\gamma)\,
	\frac{192 \,\pi\, \gn^2\, M}{\rs^3}
	\left[ 2\, \frac{\rs}{r} + \ln\! \left( \frac{r-\rs}{r+\rs} \right)
	\right] 
	+ \frac{C_1}{r}
	+ C_2
	+\mathcal{O}(\gn^3)
	\nonumber
	\\
	\delta g_{rr}^{\rm ext}
	&=
	(\alpha - \gamma) \,\frac{384\, \pi\, \gn^2\, M}{r \,(r^2 - \rs^2)}
	+\frac{C_1}{r}
	+ \mathcal{O}(\gn^3)
	\ ,
\end{align}
%}
where $C_i$ are integration constants which can be set to zero. This ensures asymptotic flatness and that the ADM mass is $M$. 

We work with the metric with signature $(+---)$, in the
signature $(-+++)$ case, the corrections obtain an extra minus sign. Note the two terms in large brackets, when combined, give rise to the $r^{-3}$ and $r^{-5}$ corrections mentioned in the introduction. The coefficient of this term is proportional to $\gn^2 M  \rs^{-3}$: i.e., it is a quantum gravitational effect proportional to the density of the source object. Two source objects with the same mass $M$ but different densities give rise to different metric perturbations.
\par

Now compare the result to that generated by two nested (one inside the other) dust balls with densities
\begin{equation}
\rho_i(r)
=
\rho_{0,i}\,
\Theta(R_i-r)
=
\begin{cases}
\rho_{0,i}
&
{\rm if}\ r<R_i
\\
0
&
{\rm if}\ r>R_i
\ ,
\end{cases}
\label{rho02}
\end{equation}
and masses $M_1$ and $M_2$
\begin{equation}
M_{i}
=
4\,\pi
\int_0^{R_{i}}
\rho_{0,i}\,r^2\,\d r
=
\frac{4\,\pi}{3}\,R_{i}^3\,\rho_{0,i}
\end{equation}
with $i\in\{1,2\}$, such that, e.g., $R_{1}<R_s$, $R_{2}=R_s$ and $M=M_1+M_2$. In other words, the star built from two nested dust balls has total mass equal to $M$ and the same outer radius $R_s$ as the star composed of only one component.

It is straightforward to show that a solution in general relativity exists. In the region $r\in [R_{2},\infty)$, the metric is the exterior Schwarzschild solution with mass $M$. In the region $r\in [0,R_{1})$ (the most inner one), the metric is the interior Schwarzschild solution with radius $R_{1}$ and mass $M_1+M_2(R_{1}/R_{2})^3$.  In the region $r\in [R_{1},R_{2})$,  the metric is the interior Schwarzschild solution with radius $R_{2}$ and mass $M_2$.

In general relativity, an external observer cannot differentiate a star with radius $R_s$ and mass $M$ from the star with two different components but same external radius and same total mass $M$.  However, we will show that the quantum gravitational corrections are different for the two matter distributions and there is thus a memory effect. Repeating the same calculation as in \cite{Calmet:2019eof}, using the fact that at this order in $\gn$ the equations are linearized, we find a correction 
\begin{align}
\label{eq:MetricCorr2}
	\delta g_{tt}^{\rm ext}
	&=
	(\alpha + \beta + 3\,\gamma)\,
	\frac{192 \,\pi\, \gn^2\, M_1}{R_{1}^3}
	\left[ 2\, \frac{R_{1}}{r} + \ln\! \left( \frac{r-R_{1}}{r+R_{1}} \right)
	\right] 
	\nonumber
	\\
	&\quad
	+ (\alpha + \beta + 3\,\gamma)\,
	\frac{192 \,\pi\, \gn^2\, M_2}{R_{2}^3}
	\left[ 2\, \frac{R_{2}}{r} + \ln\! \left( \frac{r-R_{2}}{r+R_{2}} \right)
	\right] 
	+\mathcal{O}(\gn^3)
	\nonumber
	\\
	\delta g_{rr}^{\rm ext}
	&=
	(\alpha - \gamma) \,\frac{384\, \pi\, \gn^2\, M_1}{r \,(r^2 - R_{s,1}^2)}
	+	(\alpha - \gamma) \,\frac{384\, \pi\, \gn^2\, M_2}{r \,(r^2 - R_{s,2}^2)}
	+ \mathcal{O}(\gn^3)
	\ .
\end{align}
While the classical part of the metric cannot distinguish between the one ball of dust with mass $M$ and two concentric dust balls with masses $M_1, M_2$ and $M_1+M_2=M$, the quantum gravitational corrections depend on the matter distribution of the nested balls.  

For the one-layer star we obtain
\begin{eqnarray}
g_{tt}&=&1-\frac{2 G_N M}{r} \\ && \nonumber -128 \pi^2(\alpha+\beta+3\gamma)  \frac{l_p^2}{r^2}\left[ \frac{G_N M}{r} \left (1+\frac{3 R_s^2}{5 r^2} +{\cal O} (R_s/r)^4 \right) +{\cal O}(G_N M/r)^2\right] \\ && \nonumber + ~ {\cal O} (l_p/r)^4\ ,
\end{eqnarray}
where $l_p=\sqrt{\hbar G}$ is the Planck length, and for two layers we obtain
\begin{eqnarray}
g_{tt}&=&1-\frac{2 G_N M}{r} \\ && \nonumber -128 \pi^2(\alpha+\beta+3\gamma)  \frac{l_p^2}{r^2}\left[ \frac{G_N M}{r} \left (1+\frac{3 (M_1 R_1^2+M_2 R_s^2)}{5 M r^2} +{\cal O} (R_s/r)^4 \right) +{\cal O}(G_N M/r)^2\right] \\ && \nonumber
+ ~ {\cal O} (l_p/r)^4.
\end{eqnarray}
Clearly, the quantum gravitational corrections are different for the two stars. Here we made explicit the different expansion parameters. The series in $l_p/r$ reflects the truncation of the effective action at second order in curvature. 
The series in $G_N M/r$ is due to the linearization of the field equations and the expansion in  $R_s/r$ corresponds to the asymptotic limit. In this limit we see that potentials generated by the two stars are composition dependent at order $r^{-5}$.

In this case we have considered a two-layered star and shown that the result can differ from a single-layered star. However, the above argument can easily be extended to show that any $n$- and $m$-layered stars with $n\neq m$ can be distinguished by an outside observer due to quantum gravitational effects, although their classical external gravity fields are identical. The quantum memory effect leads to hairy stars.
\par

To extend the above discussion, consider two homogeneous stars both with initial mass $M_i$ and radius $R_i$. We assume that at a certain time both stars run out of fuel and collapse towards a new equilibrium state with mass $M_f$ and radius $R_f$. Let us furthermore assume that the first star remains homogeneous, while the second collapses to a two-layered state as described above. The initial configurations are gravitationally indistinguishable in terms of classical effects. Moreover, due to Birkhoff's theorem the two final states are classically indistinguishable. However, due the quantum gravitational memory effect the two final states are distinguishable at the quantum level. 

While earlier we assumed a time-independent static star, we could consider a collapsing dust ball which can form a black hole. We introduce time-dependence via the radius of the star $R_s(t)$. For a distant observer, $r\gg R_s(t)$ at all times, we can expand the correction to the metric in Eq. (\ref{eq:MetricCorr}), and it seems likely that the $r^{-5}$ dependence remains during the totality of the collapse. 

Eventually, $R_s(t)$ will reach $2 G_N M$ and a closed trapped surface will form indicating the formation of a black hole. An observer could in principle measure the coefficient of the $r^{-5}$ correction to the metric. This correction contains information about the matter distribution that collapsed and could thus enable the observer to differentiate between black holes formed by different matter distributions. 

The $r^{-5}$ correction shifts the location of the horizon slightly and modifies the metric near the horizon. This presumably has an effect on Hawking radiation. A fully quantum mechanical treatment of the metric $g$, as opposed to the semiclassical perturbation analysis above, would yield the detailed quantum state of the graviton field (analogous to (\ref{coherent})) in place of the $r^{-5}$ correction we obtained \cite{footnote1}. 

We find that quantum gravity produces a new kind of hair on black holes. While the corrections described in this section may be very small, with limited experimental consequences, they can have dramatic consequences for black holes information \cite{Calmet:2021cip}.

\section{Conclusions: holography and black hole information}

The existence of a one to one map between the quantum states of compact matter sources and of their asymptotic gravitational fields is clearly suggestive of holography and area bounds on entropy. We emphasize that the appearance of the charge or energy in results like (\ref{coherent}) originates in Gauss law constraints which play an important role in the quantization of gauge theories and gravity. The recovery of bulk information from asymptotic gravitational fields at the boundary is also discussed in \cite{Marolf:2008mf,Laddha:2020kvp,Chowdhury:2020hse,Chowdhury:2021nxw}.  

In a fully quantum mechanical treatment the evolution of the matter source cannot be considered independently from that of its gravitational field. This contrasts sharply with the usual approximation of a fixed spacetime background in which matter fields evolve. For example, Hawking radiation from a black hole is computed in this approximation, whereas our analysis shows that a precise treatment (e.g., one which hopes to examine the unitarity of black hole evaporation) must consider that the metric outside the horizon depends on the state of the interior. The evaporation process takes the form
\begin{equation}
\vert B_0, g_0 \rangle \, \rightarrow \, \vert B_1,g_1, \gamma_1 \rangle \, \rightarrow \,  \vert B_2,g_2, \gamma_2,\gamma_1 \rangle \, \rightarrow \,  \vert B_3,g_3, \gamma_3, \gamma_2,\gamma_1 \rangle  \cdots     
\end{equation}
where $B$ is the black hole internal state, $g$ the quantum state of the (external) graviton field or metric, and $\gamma$ the emitted radiation which originates at the horizon. The radiation state $\gamma_{n+1}$ depends on the metric state $g_n$, and each $g_n$ depends on, and is entangled with, $B_n$. From this perspective it is clear that the Hawking radiation state is connected to the internal state of the black hole. 

We can give some idea of the complexity of this process through the following schematic description. Consider the semiclassical superposition state in (\ref{super}), 
\begin{equation}
\psi_g (S) = \sum_n c_n \psi_g (E_n)   ~~,
\end{equation}
and suppose that each graviton state $\psi_g (E_n)$ (describing the exterior metric) has amplitude $\alpha(E_n, \Delta)$ to produce a Hawking radiation quantum $\gamma$ with energy $\Delta$ (represented by $\gamma ( \Delta )$ in the wavefunction below). Then the exterior state evolves to
\begin{equation}
\psi  \approx \sum_n c_n \left[ \psi_g (E_n) ~+~ \alpha (E_n, \Delta) \psi_g (E_n - \Delta) \gamma ( \Delta ) ~+~ \cdots \right]
\end{equation}
The state after radiation emission (from second term in the sum, above) is a different semiclassical state constructed from $\psi_g$ corresponding to energies shifted by $\Delta$. Through $\alpha (E_n, \Delta)$ and $\psi_g (E_n - \Delta)$ the detailed form of this quantum state depends on the emitted radiation, including on quantum numbers we have suppressed such as momentum, spin, charge, etc. 
Even if the deviation of $\alpha (E_n, \Delta)$ from the semiclassical amplitude is exponentially small, the aggregate effect on the process of evaporation could be significant. It is plausible that each initial black hole state, specified by coefficients $c_n$, evolves into a different final quantum state -- i.e., the evolution is unitary. 

For each history of radiation quanta $\{ \gamma_1, \gamma_2, \cdots , \gamma_n \}$ there is a corresponding quantum spacetime $\{ g_1, g_2, \cdots , g_n \}$. A black hole with entropy $A$ can produce $\sim \exp A$ distinct evaporation states and corresponding quantum spacetimes. Schr\"odinger evolution of the initial state will produce a superposition of these radiation states and spacetimes \cite{Buniy:2020dux,Hsu:2009ve,Hsu:2010sb}. It has been conjectured that black hole evaporation is unitary when all of these branches of the wavefunction are taken into account \cite{Hsu:2013cw,Hsu:2013fra,Bao:2017who,Raju:2020smc}.

\bigskip

{\it Acknowledgments:}
The work of X.C. is supported in part  by the Science and Technology Facilities Council (grants numbers ST/T00102X/1, ST/T006048/1 and ST/S002227/1). The work of R.C. is partially supported by the INFN grant FLAG, it has also been carried out in the framework of activities of the National Group of Mathematical Physics (GNFM, INdAM).

\bigskip 

%\newpage


\begin{thebibliography}{10}
\bibitem{FN} Some details of the gedanken construction: 1. Place the $N$ copies of the system $S$ at distances $r$ apart, where $r$ is much larger than the size of $S$. We can stabilize the copies against their mutual gravitational attraction by assuming a repulsive force mediated by a boson with mass $\sim 1/r$. This finite range interaction is negligible at asymptotically large distances. 2. Consider the graviton field at distances $R$ much larger than  $N^{1/3}~ r$ (i.e., far from all of the matter sources). 3. Take the limits $R, N, r ~\rightarrow~ \infty$ such that the leading contribution to the Newtonian potential at large $R$ is given by the total energy
$E=N\, E_n$, up to corrections that can be made as small as desired.
\end{thebibliography}
\end{document}